\documentclass[prl,twocolumn,superscriptaddress]{revtex4}

\usepackage{amsmath,amsfonts}

\usepackage{graphicx}
\usepackage{times}

\newcommand{\R}{{\mathbb{R}}}
\newcommand{\C}{{\mathbb{C}}}
\newcommand{\Q}{{\mathbb{Q}}}
\newcommand{\U}{{\mathbb{U}}}

\begin{document}



\title{Large Fourier transforms never exactly realized by braiding conformal blocks}

\author{Michael H. Freedman}
\affiliation{Microsoft Project Q, Kavli Institute for Theoretical
Physics, University of California, Santa Barbara, CA 93106-4030}

\author{Zhenghan Wang}
\affiliation{Microsoft Project Q, Kavli Institute for Thereotical
Physics, University of California, Santa Barbara, CA 93106-4030}
\affiliation{Department of Mathematics, Indiana University,
Bloomington, IN 47405}


\begin{abstract}

Fourier transform is an essential ingredient in Shor's factoring
algorithm.  In the standard quantum circuit model with the gate
set $\{\U(2), \textrm{CNOT}\}$, the discrete Fourier transforms
$F_N=(\omega^{ij})_{N\times N},i,j=0,1,\cdots, N-1,
\omega=e^{\frac{2\pi i}{N}}$, can be realized exactly by quantum
circuits of size $O(n^2), n=\textrm{log}N$, and so can the
discrete sine/cosine transforms. In topological quantum computing,
the simplest universal topological quantum computer is based on
the Fibonacci (2+1)-topological quantum field theory (TQFT), where
the standard quantum circuits are replaced by unitary
transformations realized by braiding conformal blocks. We report
here that the large Fourier transforms $F_N$ and the discrete
sine/cosine transforms can never be realized exactly by braiding
conformal blocks for a fixed TQFT. It follows that approximation
is unavoidable in implementation of the Fourier transforms by
braiding conformal blocks.

\end{abstract}

\maketitle


\paragraph{Introduction.}

The simplest topological model for quantum computing which can
approximate any quantum circuit efficiently by braiding conformal
blocks is based on the Fibonacci topological quantum field theory
(TQFT) \cite{FLW02}. The corresponding conformal field theories
(CFTs) for the Fibonacci TQFT include the level=$1$ WZW $G_2$ CFT.
TQFTs are low energy effective theories for topological phases of
matter such as fractional quantum Hall (FQH) liquids, where
quasi-particles can be anyons, even non-abelions theoretically. We
will use the term anyon loosely here to include also non-abelions.
On theoretical and numerical grounds it is believed that the
Fibonnaci TQFT is an essential part of an effective theory for the
FQH liquids at filling fraction $\nu=12/5$ \cite{RR} \cite{Xia04}.
Moore and Read proposed that the ground state wavefunctions for
anyons localized at fixed positions are given by the conformal
blocks of the corresponding conformal field theory \cite{Moore91}.
Thus quantum gates in topological quantum computers are the
braiding matrices of the conformal blocks, which are also the
braiding statistics of anyons.

A decade ago, Shor discovered the polynomial-time quantum
algorithm for factoring integers.   A key component of Shor's
algorithm is the application of the discrete Fourier transforms
$F_N$.  It is known that the Fibonacci topological quantum
computer can simulate Shor's algorithm efficiently, but the
simulation requires approximations of the Fourier transforms
\cite{FLW02}. In this paper we present a \lq\lq no-go" theorem by
showing that approximation is unavoidable. Closely related to the
Fourier transforms are the discrete sine/cosine transforms which
are also useful for signal processing. Our discussion for Fourier
transforms applies equally to those transforms.

As the prospect of a topological quantum computer has attracted
increased attention, examination of the programming and compiling
issues attendant to this design has begun \cite{B05}.  Even an
accurate ($10^{-5}$) NOT gate requires several hundred elementary
braids according to the known approximation scheme \cite{NC00}.
Audiences seeing such compilations always ask, \lq\lq yes, but
isn't there a better way? Can't the arithmetic properties of
Fibonacci anyons be matched to the number theory of factoring?"
While efficient factoring is still a theoretical possibility, we
show no arithmetic wizardry will create the all-important Fourier,
sine or cosine transforms inside TQFTs.

A TQFT has a finite label set $L=\{a,b,c,\cdots \}$, which
physically represents the anyon types in the theory. Then
 a TQFT is a consistent rule to assign each 2-dimensional oriented compact
 space $\Sigma$ a vector space $V(\Sigma)$,
and each cobordism $(M,\Sigma_1, \Sigma_2)$ a linear map
$Z(M,\Sigma_1, \Sigma_2): V(\Sigma_1)\longrightarrow V(\Sigma_2)$.
In particular, a projective representation of the mapping class
group $\mathbb{M}(\Sigma)$ on $V(\Sigma)$. When $\Sigma$ has
boundaries, the boundaries will be labelled by anyons.

A TQFT is unitary if each vector space $V(\Sigma)$ has a positive
definite Hermitian inner product $<\cdot , \cdot>_{\Sigma}$
satisfying the following conditions:

1): The Hermitian inner product is multiplicative with respect to
disjoint union of surfaces, and the inner product on
$V(\emptyset)$ for the empty surface $\emptyset$ is 1.

2): The Hermitian inner product is natural with respect to the
mapping class group action.

3): For any cobordism $(M,\Sigma_1, \Sigma_2)$ and any $x\in
V(\Sigma_1)$ and $y\in V(\Sigma_2)$, we have
$$
<Z(M,\Sigma_1,\Sigma_2)(x),y>_{\Sigma_2}=<x,Z(\bar{M},\Sigma_2,\Sigma_1)(y)>_{\Sigma_1}
.$$

Theses conditions imply that the projective representations of the
mapping class groups are unitary.  Furthermore, according to
\cite{T} for any TQFT and any surface $\Sigma$ (if $\partial
\Sigma\neq \emptyset$, then $\partial \Sigma$ should be labelled)
a spanning set for $V(\Sigma)$ is obtained by the functor $V$
applied to 3-manifolds $M$ containing a labelled trivalent graph
with $\partial M=\Sigma$.  Thus for any $x, y\in V(\Sigma)$ with
$x=Z(M)$, we have $<Z(M),y>_{\Sigma}=<Z(M,\emptyset,
\partial M)(1), y>_{\Sigma}=<1, Z(\bar{M}, \partial M,
\emptyset)(y)>_{\emptyset}$.  It follows from this identity that
any Hermitian structure obeying 1)-3) above is determined by the
operators $Z(\bar{M},\partial M,\emptyset)$, hence unique.  We can
use the gluing axiom to reduce the computation of the Hermitian
inner products for all surfaces to the computation for annuli and
pairs of pants. It follows that if all the quantum dimensions of
an Hermitian TQFT are positive, and the Hermitian products on all
pairs of pants are positive definite, then the TQFT is unitary.

\paragraph{F-matrices}

Given a unitary TQFT and a 4-punctured sphere $S^2_{a,b,c,d}$,
where the 4 punctures are labelled by anyons of types $a,b,c,d$.
The 4-punctured sphere can be divided into two pairs of
pants(=3-punctured spheres) in two different ways.  In FIG.1, the
4-punctured sphere is the boundary of a thickened neighborhood of
the graph in either side, and the two graphs encode the two
different pants-decompositions of the 4-punctured sphere.  The
F-move is just the change of the two pants-decompositions.

\begin{figure}[t0]

\centerline{\includegraphics[width=3.45in]{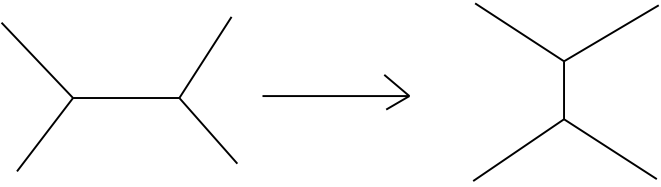}}

\caption{F-moves: the free ends are labelled by anyons}

\label{fig:F-move}

\end{figure}

By the axioms of a TQFT, each pants decomposition of
$S^2_{a,b,c,d}$ determines an orthonormal basis of
$V(S^2_{a,b,c,d})$.  Therefore the F-move gives rise to a change
of orthonormal bases of the same Hilbert space $V(S^2_{a,b,c,d})$,
hence induces a unitary matrix $F_{a,b,c,d}$, which is called the
F-matrix.

From the definition the F-matrices are unitary, but it is not
obvious that the entries of the F-matrices are always algebraic.
One of our goals is to show that the entire unitary structure,
including the F-matrices, is compatible with algebraic choices for
all unitary TQFTs.  The difficulty lies in the choices of the
F-matrices as they are basis dependent.  The obvious \lq\lq
solution": solve all complex equations in a TQFT such as the
pentagon and hexagon equations for their real and imagine parts
independently plus the unitarity constraints for the F-matrices,
is not sufficient for the compatibility as the condition of being
purely real or purely imaginary is not algebraic. Our approach
instead is to satisfy the algebraic conditions first for the
F-matrices with certain normalization and then to deduce unitarity
from the normalization.

\paragraph{Fibonacci TQFT}

First we recall the data for the Fibonacci TQFT, our chief
example. There is only one non-trivial anyon type $\tau$ in the
theory.  We will also use $\tau$ to denote the golden ratio
$\tau=\frac{1+\sqrt{5}}{2}$, and no confusions should arise.

There are two unitary TQFTs with anyon types $\{1,\tau\}$ and the
fusion rule: $\tau \otimes \tau=1\oplus \tau$.  One is the mirror
(or parity reversed) theory of the other.  We list the data for
one theory and refer to the resulting theory as the Fibonacci
TQFT. The data for the other theory is obtained by complex
conjugate all the data below.

Anyon types: $\{1,\tau\}$

Fusion rule: $1\otimes \tau=\tau\otimes 1=\tau, \tau \otimes
\tau=1\oplus \tau$

Quantum dimensions: $\{1,\tau\}$

Twists: $\theta_1=1, \theta_{\tau}=e^{\frac{4\pi i}{5}}$

Braidings: $R_1^{\tau \tau}=e^{\frac{4\pi i}{5}},
R_{\tau}^{\tau\tau}=e^{\frac{7\pi i}{5}}$

S-matrices: $ S_1=\frac{1}{\sqrt{2+\tau}} \begin{pmatrix} 1 & \tau \\
\tau & -1
\end{pmatrix}, S_{\tau}=(e^{\frac{3\pi i}{10}})$

Topological degeneracy:

Let $\Sigma_{g,n}$ be the genus=$g$ oriented surface with $n$
boundaries labelled by $\tau$, then $dim
V(\Sigma_{g,n})=\frac{\tau^n+(-1)^n
\tau^{2-2g-n}}{(\tau+2)^{1-g}}$.

Topological inner product:

The Hilbert space $V(\Sigma_{g,n})$ is spanned by labelled
uni-trivalent graphs $\{G\}$ in a bounding handlebody $H_{g,n}$
(for simplicity we ignore the framing subtlety.)  Given two
vectors in $v, w\in V(\Sigma_{g,n})$ represented by two graphs
$G_v, G_w$, then the inner product of $v,w$ is the topological
invariant of the 3-manifold $M$ with a trivalent graph $G$ inside
obtained from doubling the handlebodies and uni-trivalent graphs
$G_v, G_w$: glue the orientation reversed handlebody containing
$G_v$ with the handlebody containing $G_w$ by the identity map on
their boundaries.

Conformal block basis (FIG.2):

\begin{figure}[t1]

\centerline{\includegraphics[width=3.45in]{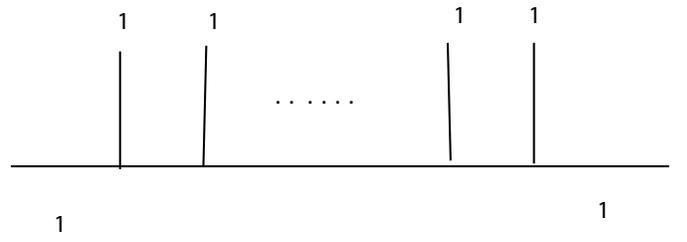}}

\caption{The basis is in one-one correspondence to admissible
labellings of the internal edges with $1$ or $\tau$ subject to the
fusion rules at each trivalent vertex. In all figures, label 0
represents type 1, and label 1 represents $\tau$.}

\label{fig:cbasis}

\end{figure}

F-matrices:  $ F=\begin{pmatrix} \tau^{-1} & {\tau}^{-1/2} \\
{\tau}^{-1/2} & -\tau^{-1}
\end{pmatrix}$

The braiding of two anyons in a conformal block basis state is
represented by the following graph (FIG. 3):

\begin{figure}[t2]

\centerline{\includegraphics[width=3.45in]{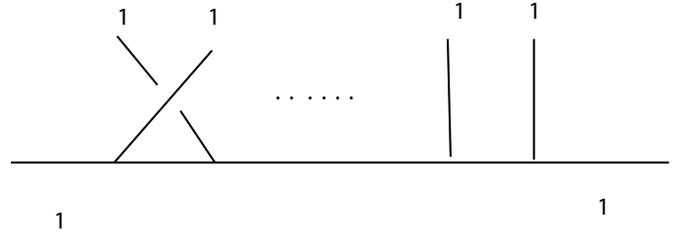}}

\caption{The braiding is obtained by stacking the braid on top of
a conformal block basis, and the braining matrix is computed by
using the graphical calculus.}

\label{fig:brading}

\end{figure}

To find the matrix elements, we form the inner products of this
braided basis with all basis.  The topological inner product in
the conformal block basis is given by flipping over the first
argument and stacking on top of the second argument.  Hence the
matrix element is an invariant of a trivalent graph with certain
braidings.  Now we observe that the invariant of any such graph is
a complex number in the number field
 $\Q(\sqrt{\tau},\xi_{20})$, where $\xi_{20}=e^{2\pi i/20}$ and
 $\Q(\sqrt{\tau},\xi_{20})$, an example of a number field,
  consists of all complex numbers which
 are rational polynomials in $\sqrt{\tau}, \xi_{20}$ with integer
 coefficients.  Therefore, we have:

{\it Observation:} All matrix entries of the braiding matrices
with the above choices of data for the Fibonacci TQFT lie inside
the number field $\Q(\sqrt{\tau},\xi_{20})$ whose Galois group is
the non-abelian dihedral group $D_4$.  Furthermore, only
1,2,4,5,10,20-th roots of unity exist in
$\Q(\sqrt{\tau},\xi_{20})$.

 We will now see that there are only finitely
 many roots of unity in $\Q(\sqrt{\tau},\xi_{20})$.  But to
 realize all the discrete Fourier transforms $F_N$, we need
 infinitely many root of unity, therefore discrete Fourier
 transforms $F_N$ for large $N$ cannot be realized exactly by
 braiding conformal blocks.  The roots of unity in
 $\Q(\sqrt{\tau},\xi_{20})$ determine which Fourier transform can be
 potentially realized by braiding conformal blocks in the
 Fibonacci TQFT.  The notation below is
clarified in the following section.  Notice that
$[\Q(\sqrt{\tau},\xi_{20}): \Q]=
 [\Q(\sqrt{\tau},\xi_{20}):\Q(\xi_{20})][\Q(\xi_{20}):\Q].$
 For a primitive m-th root of unity $\xi_m$,
 $[\Q(\xi_{m}):\Q]=\phi(m)$, where $\phi(m)$ is the
Euler function whose value is the number of integers from $1$ to
$m-1$ that is relatively prime to $m$.  Since $\tau \in
\Q(\xi_{20})$, so $[\Q(\sqrt{\tau},\xi_{20}):\Q(\xi_{20})]=2$, and
$[\Q(\xi_{20}):\Q]=\phi(20)=8$.  Hence we have
$[\Q(\sqrt{\tau},\xi_{20}):\Q]=16$.  It follows that there are
only finitely many roots of unity in $\Q(\sqrt{\tau},\xi_{20})$
since there are only finitely $m$ such that $\phi(m)\leq 16$.

 Another consequence of this observation is that the Fibonacci
 TQFT cannot be realized in an abelian extension of $\Q$ because
 the Galois group of $\Q(\sqrt{\tau},\xi_{20})$ is non-abelian.
 It suffices to show that the Galois group of $\Q(\sqrt{\tau}, i) \subset
 \Q(\sqrt{\tau},\xi_{20})$ is
 non-abelian, which is the same as the Galois group for the minimal polynomial
 of $\sqrt{\tau}$ $f(x)=x^4-x^2-1$.  To determine the Galois group
 of $f(x)$, we use the following fact: let $g(x)=x^4+ax^2+b\in \Q(x)$
 be irreducible with Galois group $G$. If neither $b$ nor
 $b(a^2-4b)$ is a square in $\Q$, then the Galois group of $g(x)$
 is the non-abelian dihedral group $D_4$.  For a proof, see
Proposition
 4.11 of \cite{Hung} on Page 273 and Ex. 9 on page 277.  Now it is
 obvious that the Galois group of $f(x)$ is $D_4$.

Finally let us determine all the possible roots of unity in
$\Q(\sqrt{\tau},\xi_{20})$.  If a primitive m-th root of unity
$\xi_m$ is in $\Q(\sqrt{\tau},\xi_{20})$, then $\phi(m)$ is a
factor of $16$ because
$[\Q(\sqrt{\tau},\xi_{20}):\Q]=[\Q(\sqrt{\tau},\xi_{20}):\Q(\xi_{m})]
[\Q(\xi_{m}):\Q]=16.$  If $m$ is relatively prime to $20$ and
$\geq 7$, then $\xi_{20m}$ would be in $\Q(\sqrt{\tau},\xi_{20})$.
But $\phi(20m)=8\phi(m)>16$, a contradiction.  It follows that if
$\xi_m \in \Q(\sqrt{\tau},\xi_{20})$, then $m$ is of the form
$2^k\cdot 3\cdot 5$ for possibly $k=1,2,3$.  But first $3$ cannot
be a factor $m$ because otherwise, $\xi_{60}\in
\Q(\sqrt{\tau},\xi_{20})$. Since $\Q(\xi_{60})$ would be a
subfield of $\Q(\sqrt{\tau},\xi_{20})$ which are both degree $16$
extension of $Q$, we will have
$\Q(\xi_{60})=\Q(\sqrt{\tau},\xi_{20})$.  But this is impossible
since the Galois group of $\Q(\xi_{60})$ is abelian, while the
Galois group of $\Q(\sqrt{\tau},\xi_{20})$ is non-abelian. Exactly
the same argument will rule out $k=3$ with $\xi_{40}$ replacing
$\xi_{60}$. So the only possible primitive roots of unity in
$\Q(\sqrt{\tau},\xi_{20})$ are $\xi_{m}, m=1,2,4,5,10,20$, and
their powers.

Using the relation $\frac{p_{+}}{D}=e^{2 \pi i \cdot c/8 }$, we
deduce that the central charges of the corresponding CFTs are
$c=14/5$ mod 8, which is realized by the level=$1$ $G_2$ CFT.
 Because the central charges $c\neq 0$, we have to either work with
projective representations rather than linear representations of
the mapping class groups or work with some central extension of
the mapping class groups for extended surfaces.  For the torus
case, the projective representation can always be lifted to a
linear representation as follows: direct computation shows that
$(st)^3=\frac{p_{+}}{D} s^2$, so if we set $\tilde{t}=t
(\frac{p_{+}}{D})^{-\frac{1}{3}}$, then $(s\tilde{t})^3=s^2$.  It
has been shown that a 3rd-root of unity of $\frac{p_{+}}{D}$ is
sufficient to lift all projective representations of the mapping
class groups to linear representations of the extended mapping
class groups \cite{T}.  Hence there are at least three different
normalizations for a given TQFT, which lead to successively larger
number fields:

1).  Arbitrary choice for the F-matrices and projective
representations for the mapping class groups

2).  Unitary normalization for the F-matrices and projective
representations for the mapping class groups

3). Unitary normalization for the F-matrices and linear
representations for the extended mapping class groups.

For the Fibonacci TQFT, with normalization 1), the Fibonacci TQFT
can be described in $\Q(\xi_{20})$; with normalization 2),
$\Q(\sqrt{\tau},\xi_{20})$; with normalization 3)
$\Q(\sqrt{\tau},\xi_{60})$.  Note this field contains $\xi_m$ for
all $m|60$ by an argument similar to the one above.

\paragraph{Unitary TQFTs}

Let $\Q$ be the field of the rational numbers (a field here is not
in the sense field theory in physics, but as in number theory.  A
field is a generalization of the number systems $\Q,\R,\C$.)  A
number field is a finitely dimensional vector space $K$ over $\Q$
which is a field: a vector space with a compatible multiplication.
The field $K$ is called an extension field of $\Q$, and the
dimension of $K$ as a vector space over $\Q$ is called the degree
of the extension, denoted by $[K:\Q]$. Given a complex number $x$,
$\Q(x)$ is the field of all complex numbers of the form
$p(x)/q(x)$, where $p(x),q(x)$ are polynomials in $x$ with
coefficients in $\Q$ and $q(x)\neq 0$.  For example,
$\Q(\sqrt{\tau})$ is a degree=$4$ extension of $\Q$.  Fields can
be extended repeatedly as follows: let $K$ be an extension of $\Q$
and $y$ a complex number, then $K(y)$ is the field of all complex
numbers of the form $p(y)/q(y)$, where $p(y),q(y)$ are polynomials
in $y$ with coefficients in $K$ and $q(y)\neq 0$. Given two
complex numbers $x,y$, the number field $\Q(x,y)$ is the extension
of first $\Q$ to $\Q(x)=K$ or $\Q(y)=K$, then $K$ to $K(y)$ or
$K(x)$, which both are $\Q(x,y)$. The degree of the extension is
$[\Q(x,y):\Q]=[\Q(x,y):\Q(x)][\Q(x):\Q]=[\Q(x,y):\Q(y)][\Q(y):\Q]$.

\paragraph{Main Results:}

1.  Given a unitary TQFT, there is a normalization so that all the
entries of the F-matrices are in a number field K, and the
F-matrices associated to the F-moves are unitary.

2.  Each Hilbert space $V(\Sigma)$ has an orthonormal basis so
that every representation matrix of the mapping class group has
entries in the number field $K$.  Warning the Galois group
$Gal(K/\Q)$ is not necessarily abelian.

3.  Large Fourier transforms, the discrete sine/cosine transforms
cannot be realized exactly in any fixed TQFT by braiding conformal
blocks.

Parts 3 follows from Part 2 as follows. We recall that the number
of roots of unity in a number field $K$ is always finite. To see
this, the degree of the extension of $\Q(\xi_m),
\xi_m=e^{\frac{2\pi i}{m}}$ is $\phi(m)$, where $\phi(m)$ is the
Euler function. Therefore if $[K:\Q]=n$ and $\phi(m)>n$, then
$\xi_m$ cannot be in $K$ because otherwise the extension degree
will be $>n$. Similarly for $\Q(sin(2\pi /m)), \Q(cos(2 \pi /m))$.

Part 2 follows from Part 1. Given a unitary TQFT, there is a
unique way to construct compatible topological inner products for
$V(\Sigma)$'s [\cite{T}, Chapter IV, Section 10], and we need an
explicit orthonormal basis for each $V(\Sigma)$ to compute the
braiding matrices. To do this one sets up a graphical calculus so
that each matrix entry is an invariant of a certain trivalent
graph, which depends on our choices of the F-matrices and $\theta$
symbols. The theorem is reduced to the careful choices of
F-matrices and $\theta$ symbols which are compatible with the
topological inner product. The invariants of such graphs are
polynomial of certain roots of unity and $6j$ symbols
$F_{ijk}^{lmn}$. There are three kinds of contributions of roots
of unity: the braiding eigenvalues, the twists, and the higher
Frobenius-Schur indicators resulting from bending anyon
trajectories. They are in some fixed extension of $\Q$ whose
degree is determined by the fusion rules through Vafa's theorem.
The $6j$ symbols are constrained by the pentagon identities. To
have a consistent set of $6j$ symbols $F_{ijk}^{lmn}$ with
graphical calculus, it is sufficient to solve the following set of
polynomial equations (for easiness of notation we drop the
dependence on trivalent vertices):

1.  $F_{j^*i^*0}^{ijk}=\sqrt{\frac{d_k}{d_id_j}}\delta_{ijk}$

2.
$F_{kln}^{ijm}=F_{ijn^*}^{klm^*}=F_{nk^*l^*}^{mij}\sqrt{\frac{d_md_n}{d_jd_l}}$

3.
$\sum_{n}F_{kp^*n}^{mlq}F_{mns^*}^{jip}F_{lkr^*}^{js^*n}=F_{q^*kr^*}^{jip}F_{mls^*}^{riq^*}$

Any solution of this set of equations will be a consistent choice
of 6j symbols for the unitary TQFT.  Now we cite a theorem in
algebraic geometry: the solution to the polynomial equations above
is an algebraic variety over $\Q(\sqrt{d_i}), i=1,2,\cdots, R$,
where $R$ is the number of anyon types. Since this variety has at
least one point which gives rise to the TQFT, then there will be
also an algebraic point by Theorem 7 on Page 32 \cite{Lang}. It
follows that every graph invariant will be inside a fixed finite
extension of $\Q{(\sqrt{d_i})}$ and hence in a number field over
$\Q$.

The resulted graphical calculus from the solution of the the
pentagon equations with the above normalization has very nice
properties.  The conformal block basis is an orthogonal basis. The
$\theta$ symbols $\theta(a,b,c)=\sqrt{d_a d_b d_c}$, where $d_a,
d_b, d_c$ are the quantum dimensions of the anyons $a,b,c$. One
consequence of the $\theta$ symbol values is that the conformal
block basis elements have the same length, independent of the
internal labellings.  So the $F$-matrices are change of basis for
two orthonormal bases up to overall scalars, hence are unitary.

\paragraph{Approximation by Fibonacci Quantum Computer}

Since the exact realization of the Fourier transforms is
impossible in the Fibonacci TQFT, we would like to approximate
them using braiding matrices.  Given a prescribed accuracy, it
will be interesting to find the explicit approximations.   We will
only outline an approximation here.

To simulate a standard $n$-qubit quantum circuit $U_L:
(\C^2)^{\otimes n}\rightarrow (\C^2)^{\otimes n}$, we embed
$(\C^2)^{\otimes n}$ into the conformal blocks on $2n+2$ Fibonacci
anyons at fixed positions.  Since $dim(V_{2n+2})=F_{2n+2}> 2^n$
except for $n=1$, we need to choose an efficiently computable
subspace of the conformal blocks.  One way to do this is to choose
the following subspace $(\C^2)^{\otimes n}$ of $V_{2n+2}$ with the
conformal block basis (FIG. 4).

\begin{figure}[t3]

\centerline{\includegraphics[width=3.45in]{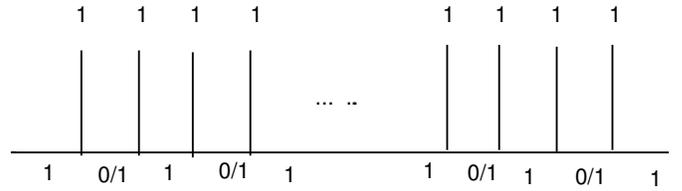}}

\caption{The edges labelled by $0$ or $1$ correspond to standard
qubits, and other basis span the non-computational subspace which
should be evolved by the identity operator ideally in the
computational process. Again 0 represents 1, and 1 represents
$\tau$}

\label{fig:computation}

\end{figure}

Then we look for a braid $b$ so that the following diagram
commutes up to the prescribed error, where $\rho(b)$ is the
braiding matrix of conformal blocks.

\begin{displaymath}
\begin{matrix} ({\mathbb{C}^2})^{\otimes n} & \longrightarrow
& V_{{2n+2}}\cr {U_L} \downarrow & & \downarrow {\rho(b)}\cr
({\mathbb{C}^2})^{\otimes n} & \longrightarrow & V_{{2n+2}}
\end{matrix}
\end{displaymath}

The standard quantum circuits for the exact realization of the
Fourier transforms are given on Page 219 of \cite{NC00}.  Given a
precision $\epsilon>0$, then one finds a braid that approximate
$F_N$ by using the approximations of the single qubit gates and
$\textrm{CNOT}$ in \cite{B05}.

\paragraph{Conclusion}

TQFTs are effective theories for topological phases of matter such
as the fractional quantum Hall liquids. Specifically, the braiding
matrices of conformal blocks are unitary transformations of the
degenerate ground states when anyons are fixed at certain
positions. Because polynomial time approximation schemes exist
\cite{NC00}, the reported obstruction to exact realization of the
Fourier transforms will not impose a fundamental physical
constraint on topological quantum computing.  However as a
practical matter there is an important distinction between
billions as opposed to millions of braid generators to factor a
large number.

The Jones braid representation(s) that we get from Fibonacci
anyons can be described as a regularized Fourier transform \lq\lq
FTB" of the braid group(s) $B_n$.  The braid generators correspond
to \lq\lq position" coordinates and the path basis of conformal
blocks is a regularized momentum basis for the group algebra of
the braid group $\C[B_n]$.  The chosen regularization consists of
passing to an appropriate semi-simple quotient, the Temperley-Lieb
algebra $\textrm{TL}^n_q=\C[B_n]/\sim, q=e^{2\pi i/5}$.  We have
shown that one cannot find the FT of large cyclic groups inside
these FTB.  The most direct application of FTB is to the
estimation of Jones polynomial evaluations \cite{FKLW} and
\cite{BFLW}.  The possibility of harnessing FTB for number
theoretic application such as factoring should be explored.

\paragraph{Acknowledgements}
This research has been supported by the NSF under grants
DMR-0130388 and DMR-0354772 (Z.W.).

\vskip -0.5cm


\end{document}